\documentclass[conference]{IEEEtran}
\IEEEoverridecommandlockouts
\pdfoutput=1
\usepackage{cite}
\usepackage{amsmath,amssymb,amsfonts}
\usepackage{algorithmic}
\usepackage{booktabs}
\usepackage{graphicx}
\graphicspath{{./figures/}}
\usepackage{textcomp}
\usepackage{tikz}
\usepackage{xcolor}
\usepackage[caption=false,font=normalsize,labelfont=sf,textfont=sf]{subfig}
\def\BibTeX{{\rm B\kern-.05em{\sc i\kern-.025em b}\kern-.08em
    T\kern-.1667em\lower.7ex\hbox{E}\kern-.125emX}}
\begin{document}

\title{Dispersion Compensation of Sinuous Antennas for Ground Penetrating Radar Applications
	\thanks{This work was supported by the U.S. Army CCDC C5ISR Center, Night Vision and Electronic Sensors Directorate, Countermine Division, Countermine Technologies Branch; the U.S. Army Research Office under Grant Number W911NF-11-1-0153; and by Sandia National Laboratories, a multimission laboratory managed and operated by National Technology and Engineering Solutions of Sandia, LLC., a wholly owned subsidiary of Honeywell International, Inc., for the U.S. Department of Energy's National Nuclear Security Administration under contract DE-NA0003525. This paper describes objective technical results and analysis. Any subjective views or opinions that might be expressed in the paper do not necessarily represent the views of the U.S. Department of Energy or the United States Government.}
}

\author{\IEEEauthorblockN{Dylan A. Crocker\textsuperscript{1,2}}
\IEEEauthorblockA{
    \textit{\textsuperscript{1}ISR EM \& Sensor Technologies} \\
    \textit{Sandia National Laboratories}\\
    Albuquerque, New Mexico, U.S.A.}
\and
\IEEEauthorblockN{Waymond R. Scott, Jr.\textsuperscript{2}}
\IEEEauthorblockA{
    \textit{\textsuperscript{2}School of Electrical and Computer Engineering} \\
    \textit{Georgia Institute of Technology}\\
    Atlanta, Georgia, U.S.A}
}

\maketitle

\begin{abstract}
Sinuous antennas exhibit many desirable properties for ground penetrating radar (GPR) applications such as ultra-wide bandwidth, polarization diversity, and a low-profile form factor. However, sinuous antennas are dispersive since the active region moves with frequency along the structure. This is an undesirable quality for pulsed-radar applications since the radiated pulse will be distorted. Such distortion may be detrimental to close-in sensing applications such as GPR. This distortion may be compensated in processing with accurately simulated or measured phase data. However, antenna performance may deviate from that simulated or measured due to the dielectric loading of the ground. In such cases, it may be desirable to employ a dispersion model based on antenna design parameters which may be optimized {\em in-situ}. Dispersion compensation models previously investigated for other antennas may be similarly applied to sinuous antennas. This paper explores the dispersive properties of the sinuous antenna and presents a simple model that may be used to compress dispersed pulses.
\end{abstract}
 

\section{Introduction}
Ground penetrating radars (GPRs) often utilize ultra-wide band (UWB) pulses to improve range resolution and thus target discrimination \cite{daniels2004ground}. Additionally, polarimetric radar techniques \cite{giuli1986polarization} have been applied to GPR in order to increase the accuracy of target classification \cite{yu2016application}. The sinuous antenna, first published in a patent by DuHamel in 1987, is an excellent candidate for such systems. The patent describes the sinuous antenna as a combination of spiral and log-periodic antenna concepts which resulted in a design capable of producing ultra-wideband radiation with polarization diversity \cite{duhamel1987dual}. Other wideband antenna designs such as quad-ridge horn \cite{blejer1992ultra}, Vivaldi \cite{pochanin2015ultra}, and resistive-vee \cite{sustman2013a} antennas provide similar capabilities. However, they require relatively large and often complex three dimensional structures in order to produce orthogonal senses of polarization. Alternatively, the sinuous antenna may be implemented as a low-profile planar structure \cite{duhamel1987dual}.

\begin{figure}[!tbp]
	\centering
	\subfloat{\includegraphics[width=0.4\columnwidth]{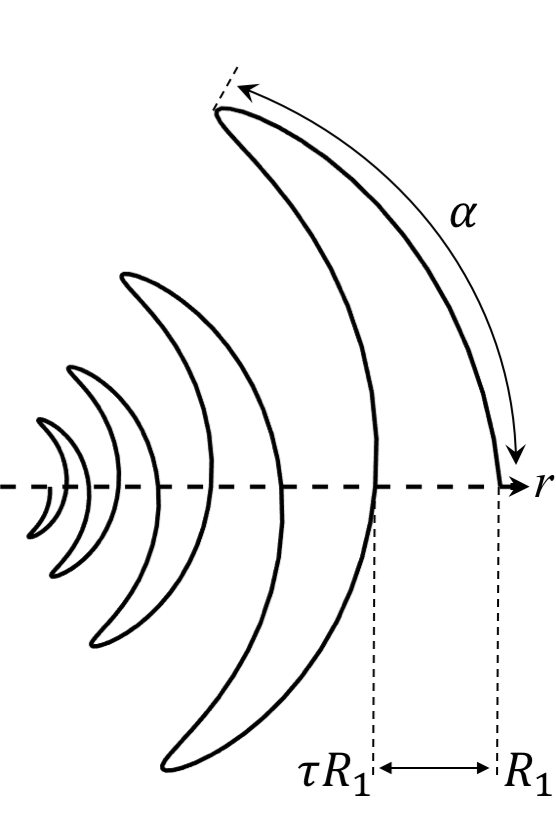}}
    \hspace{20pt}
	\subfloat{\includegraphics[width=0.4\columnwidth]{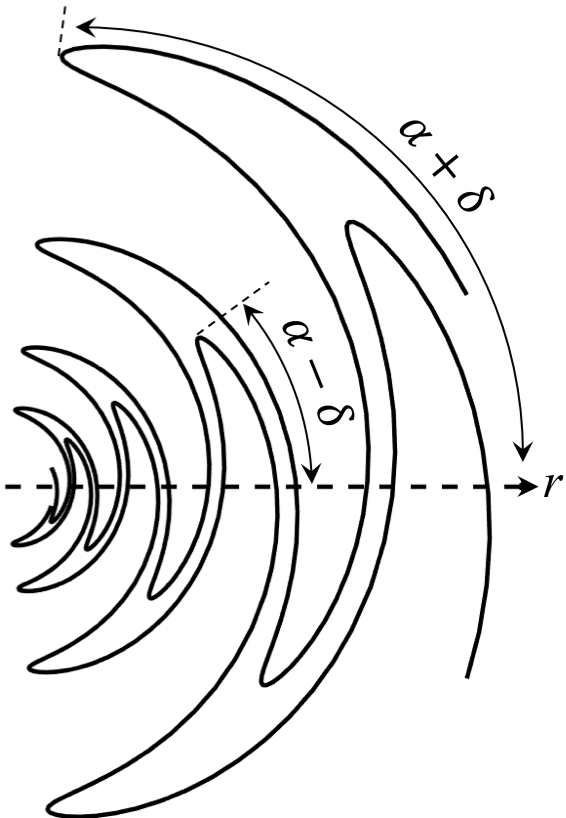}}
    \caption{Illustration of sinuous antenna design parameters: angular width $\alpha$, expansion ratio $\tau$, outermost cell radius $R_1$, and curve rotation angle $\delta$.}
	\label{fig:sinuous}
\end{figure}

Although sinuous antennas provide the desirable properties described above, they are dispersive since the active region moves on the structure with frequency. Dispersive behavior is well documented in other log-periodic antennas such as planar spirals \cite{mcfadden2009analysis}, conical spirals \cite{hertel2003on}, log-periodic dipole \cite{knop1970on, mclean2004the} and planar toothed log-periodic antennas \cite{olvera2017dispersive}. Dispersive antennas are less desirable for pulsed-radar applications as the radiated pulses become distorted in the time domain thereby reducing resolution. Dispersed pulses may be compressed during processing by applying phase corrections. Such corrections may be obtained through accurate simulation or measurement of the antenna \cite{hertel2003on}. Additionally, models based on the antenna design parameters may be used to compensate the dispersion. Such models may be desirable for GPR applications since they can be adjusted {\em in-situ} to accommodate environmental effects e.g., dielectric loading of the soil. However, for such models to remain valid, care must be taken when making sinuous antenna design decisions in order to avoid the excitation of unintended resonant modes which may result in pulse distortion not correctable with simple dispersion models \cite{crocker2019on, kang2015modification}. 

In this work, we seek to build an understanding of the dispersive nature of sinuous antennas and develop a model for its compensation. Thereby enabling GPR systems to obtain the benefits of sinuous antennas while utilizing them for UWB pulse radiation.

\section{Sinuous Antenna Dispersion}\label{sec:disp}


\begin{figure}[!tbp]
	\centering
	\includegraphics[width=0.8\columnwidth]{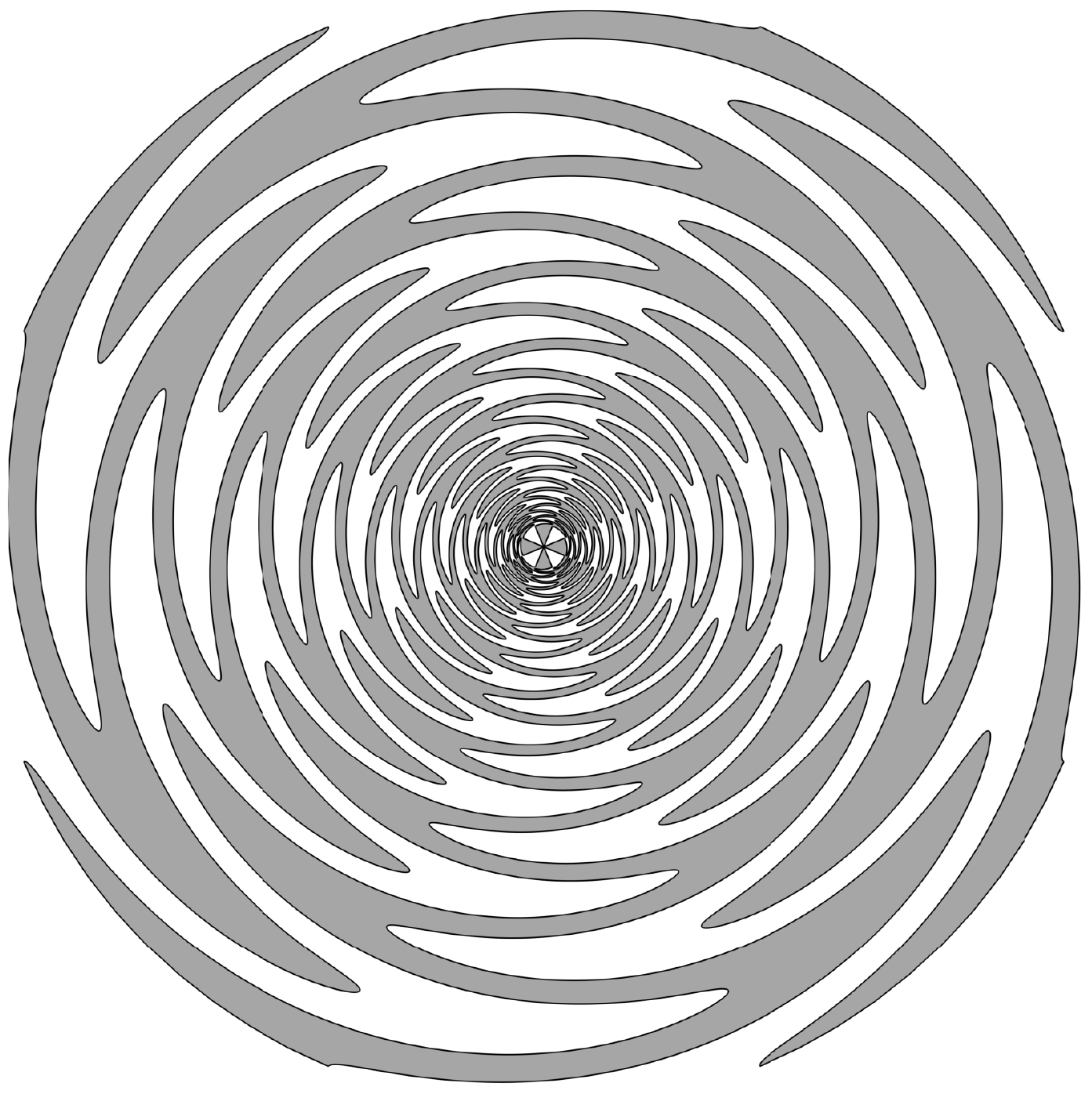}
    \caption{Sinuous antenna having parameters: $N=\text{4}$ arms, $P=\text{20}$ cells, $R_T=\text{10}$ cm, $\tau=\text{0.8547}$, $\alpha=\text{45}^\circ$, and $\delta=\text{22.5}^\circ$.}
	\label{fig:sinuous_ex}
\end{figure}

Sinuous antennas are comprised of $N$ arms each made up of $P$ cells where the curve of the $p^\text{th}$ cell is described in polar coordinates ($r$, $\phi$) by
\begin{equation}\label{eq:sin}
\phi = (-1)^{p-1}\alpha_p\sin \left( \frac{\pi \ln(r/R_p)}{\ln(\tau_p)} \right) \pm \delta,
\end{equation}
where $R_{p+1} \le r \le R_p$ \cite{duhamel1987dual}. In (1), $R_p$ controls the outer radius, $\tau_p$ the growth rate i.e., $R_{p+1}=\tau R_p$, and $\alpha_p$ the angular width of the $p^\text{th}$ cell. The curve is then rotated $\pm$ the angle  $\delta$ in order to fill out the arm as illustrated by Fig. \ref{fig:sinuous}. In this analysis, a four-arm ($N=\text{4}$) sinuous antenna is considered with $\tau$, $\alpha$, and $\delta$ constant for all cells which produces a log-periodic structure \cite{duhamel1957broadband, duhamel1987dual}. Additionally, $\delta$ is set to 22.5$^\circ$ in order to produce a self-complementary structure\footnote{
	A structure is considered self-complimentary when the metal and non-metal sections are exact replicas offset by a rotation. For sinuous antennas defined by (\ref{eq:sin}), the self-complimentary condition is met when $\delta=90^\circ/P$. The parameter $\alpha$ does not affect this condition as only $\delta$ controls the metal to non-metal ratio.
}.
The self-complementary  condition helps to ensure that the sinuous antenna's input impedance is both real and frequency independent \cite{edwards2012dual}. The antenna analyzed in this work is fed by a self-complimentary arrangement of orthogonal bow-tie elements each feeding a set of opposing sinuous arms as illustrated in Fig. \ref{fig:sinuous_ex}.

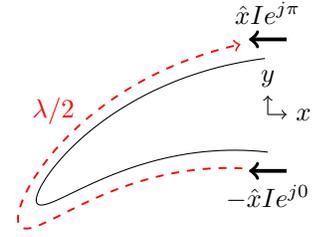
\begin{figure}
\centering
\begin{tikzpicture}[scale=5.0]
\fill [white] (-0.75,0.45) rectangle (.15,1.2);	
\draw[->] (0,.85) -- (0.05,.85) node[right] {$x$};
\draw[->] (0,.85) -- (0,.9) node[above] {$y$};	
\draw[<-, line width=0.5mm] (-0.05,1.05) -- (0.05,1.05);
\draw[->, line width=0.5mm] (0.05,.7) -- (-0.05,.7);
\draw (0,1.12) node {$\hat{x}Ie^{j\pi}$};
\draw (0,0.63) node {$-\hat{x}Ie^{j0}$};	
\draw plot [domain=0.75:1,samples=100, variable=\r] ({\r*sin(-45*sin(180*ln(\r)/ln(.75))},{\r*cos(45*sin(180*ln(\r)/ln(.75))});	
\draw (-.57,.86) node[text=red] {$\lambda/2$};
\draw[->, dashed, line width=0.25mm, color=red] plot [domain=0.71:1.04,samples=100, smooth, variable=\r] ({\r*sin(-50*sin(180*ln(\r/1.05)/ln(0.6667))},{\r*cos(50*sin(180*ln(\r/1.05)/ln(0.6667))});
\end{tikzpicture}
\caption{Illustration of sinuous antenna active region as defined in the original DuHamel patent.}
\label{fig:sinrad}
\end{figure}

Radiation from a sinuous antenna, as described in \cite{duhamel1987dual}, occurs at active regions which are formed when the length of a cell is approximately a multiple of $\lambda/2$. In this case, the current at the start and end of a cell is in phase due to the wrapping of the arm and $\lambda/2$ travel as illustrated in Fig. \ref{fig:sinrad}. These active regions move inward and outward on the antenna as the frequency increases and decreases respectively resulting in a time delay between frequencies i.e., dispersion. The dispersion increases with $\tau$ since larger values of $\tau$ result in more cells i.e., longer travel times along the arms. This may encourage GPR system designers to choose small values of $\tau$; however, larger values of $\tau$ result in better pattern uniformity \cite{duhamel1987dual, baldonero2009uwb}.

In order to investigate the dispersion, full-wave electromagnetic analysis was conducted using the CST Microwave Studio \cite{cst} time-domain solver with hardware acceleration. The sinuous antenna simulated was designed to operate from 800 MHz to 10 GHz using the following design parameters: $R_T=\text{10}$ cm, $R_{in}=\text{0.4}$ cm (bow-tie radius), $\tau=0.8547$, $\alpha=\text{45}^\circ$, and $\delta=\text{22.5}^\circ$. Note that the outer truncation radius $R_T$ (10 cm) is slightly smaller than $R_1$ in order to prevent the sharp ends produced by traditional truncation; these sharp ends are known to resonate at lower frequencies resulting in pattern distortion \cite{kang2013experimental, kang2015ends}. The antenna was simulated in free-space to simplify the analysis. Both pairs of opposing sinuous arms were terminated by an ideal port set to the theoretical impedance of 267 $\Omega$ \cite{edwards2012dual}. A single pair was then driven, with the other pair remaining matched, in order to produce linearly-polarized radiation.

\begin{figure}
	\centering{\includegraphics[width=\columnwidth]{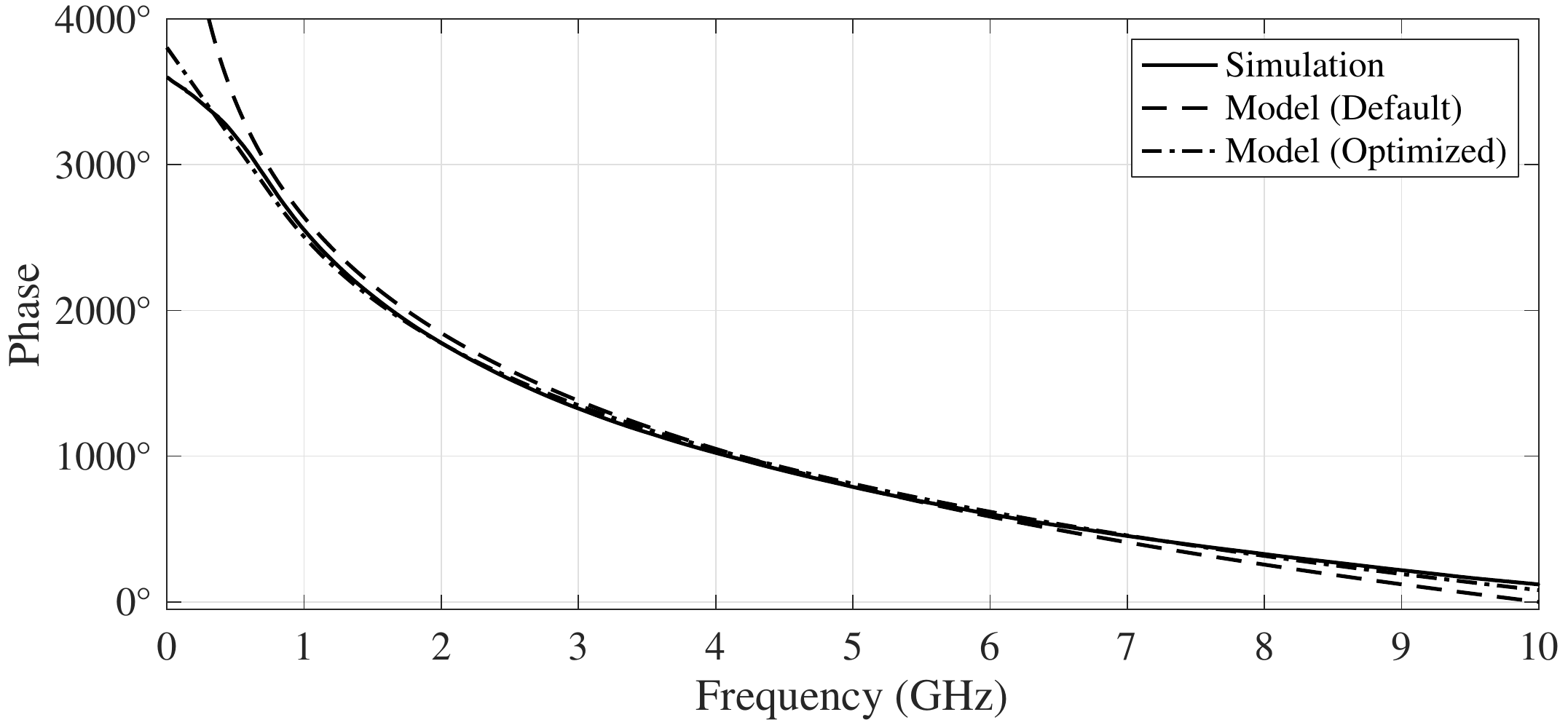}}
	\caption{Simulated and modeled phase of the radiation due to dispersion in the sinuous antenna. Phase unwrapping starts at 10 GHz.}
	\label{fig:phase}
\end{figure}

\begin{figure}
	\centering{\includegraphics[width=\columnwidth]{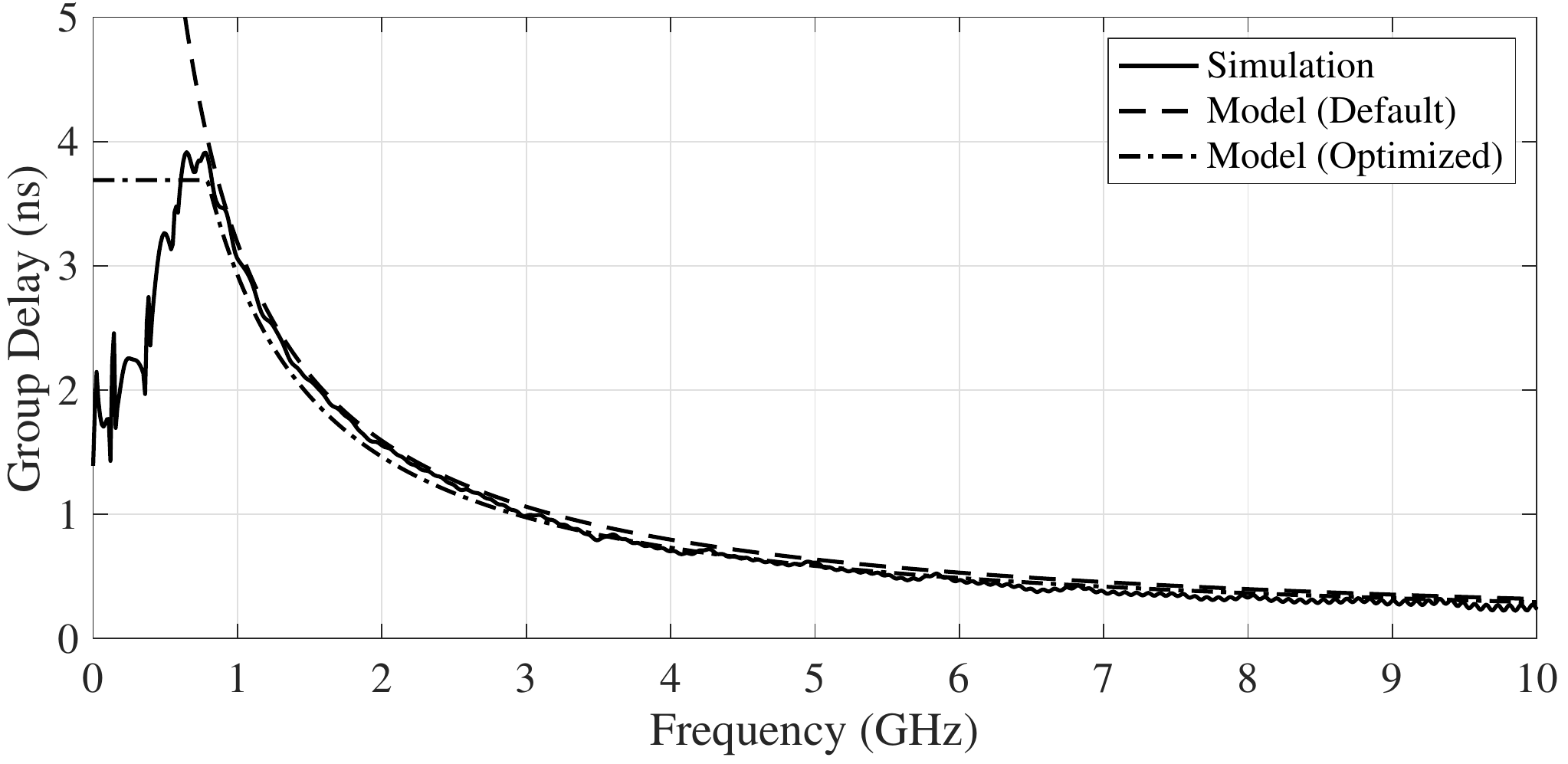}}
	\caption{Simulated and modeled group delay of the radiation due to dispersion in the sinuous antenna.}
	\label{fig:gd}
\end{figure}

The simulated co-polarized radiated fields $E^x_{sim}(r=r_p,\omega)$ were probed at a distance of $r_p=\text{2}$ m on boresight and the corresponding phase was then back-propagated to the antenna by
\begin{equation}\label{eq:back}
    \Phi^d_{sim}(\omega)=\arg\left [E_{sim}^x(r_p,\omega)\exp(jkr_p) \right ] ,
\end{equation}
leaving only the phase due to dispersion $\Phi_{sim}^d$. The phase is then unwrapped (starting with the 10 GHz sample) and shown in Fig. \ref{fig:phase}. The corresponding group delay
\begin{equation}\label{eq:gd}
    \tau^{gd}_{sim}(\omega)=-\frac{d}{d\omega}\Phi^d_{sim}(\omega),
\end{equation}
is shown in Fig. \ref{fig:gd}. As expected, lower frequencies exhibit larger delay since the corresponding active region is farther out on the antenna---where the antenna is larger. This effect is more clearly seen in the time domain.

\begin{figure}
	\centering{\includegraphics[width=\columnwidth]{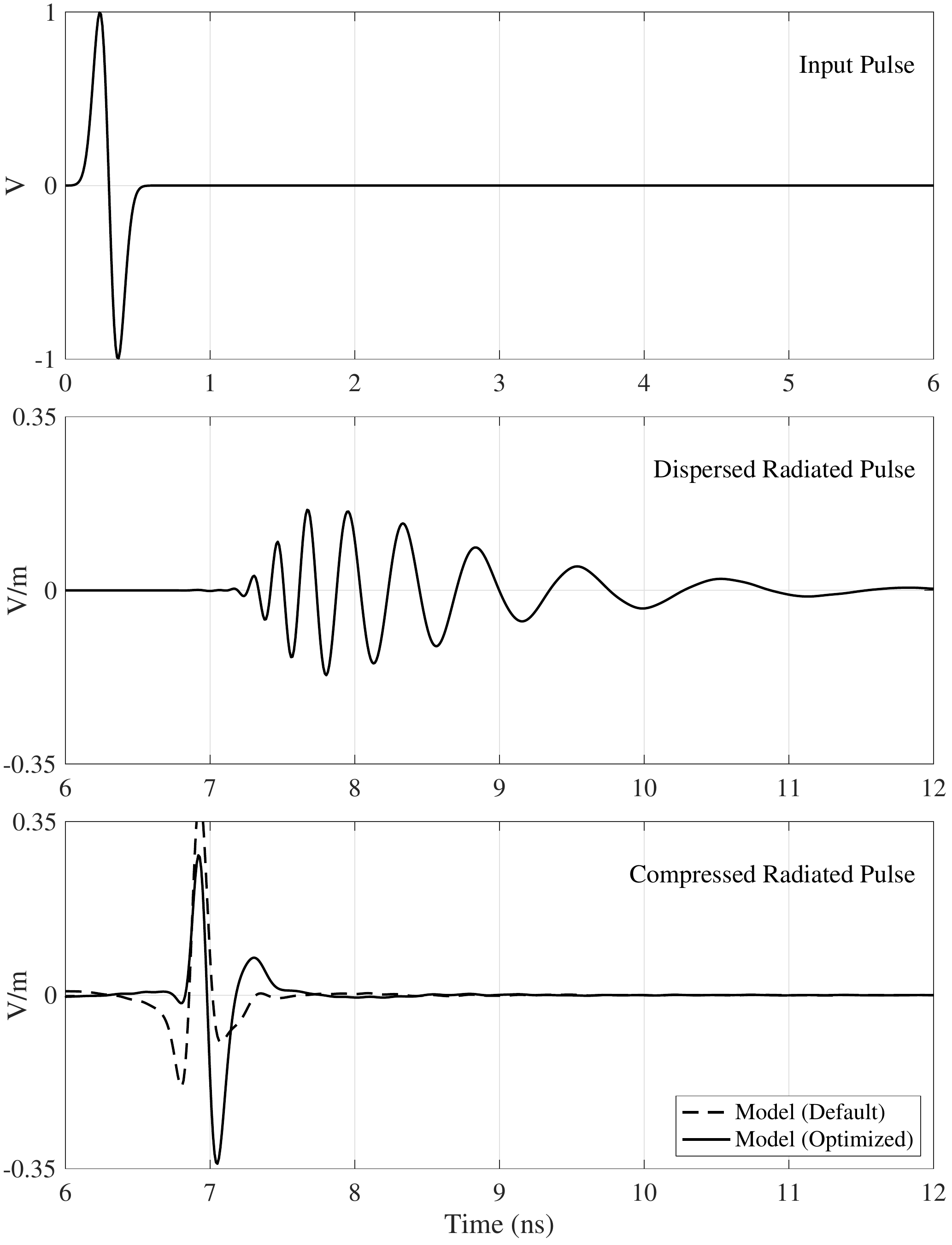}}
	\caption{Input differentiated Gaussian voltage pulse (top), corresponding dispersed radiated pulse at 2 m (middle), and the compressed radiated pulse at 2 m after the dispersion model has been applied (bottom).}
	\label{fig:pulse}
\end{figure}

The time-domain radiated pulse $E^x_{pulse}(r_p,t)$ for a given excitation $v_{pulse}(t)$ can be computed from the frequency-domain radiated fields by 
\begin{equation}\label{eq:tf}
    E_{pulse}^x(r_p,t) = \mathcal{F}^{-1}\biggl\{\mathcal{F}[v_{pulse}(t)]~\frac{E^x_{sim}(r_p,\omega)}{v_{sim}(\omega)}\biggr\},
\end{equation}
where $v_{sim}(f)$ is the frequency-domain excitation in the simulation and $\mathcal{F}$ and $\mathcal{F}^{-1}$ are the Fourier and inverse Fourier transform respectively. The pulse excitation used was a Differentiated Gaussian which is defined by
\begin{equation}\label{eq:pulse}
v_{pulse}(t) = -v_{peak}\frac{(t - \mu)}{\sigma} \exp \left [ 0.5 - \frac{(t - \mu)^2}{2 \sigma^2} \right ], 
\end{equation}
where $\mu$ represents an arbitrary time shift and the width of the pulse is controlled by $\sigma=2.3548/\omega_{BW}$. In the presented analysis, the parameters were set to $v_{peak}=\text{1 V}$ and $\omega_{BW}/2\pi=\text{6 GHz}$ resulting in peak spectral energy at 2.5 GHz. The input pulse and the corresponding radiated pulse (at $r_p=\text{2 m}$) is shown in Fig. \ref{fig:pulse}. As expected from the group delay shown in Fig. \ref{fig:gd}, the lower frequency content is delayed in time from the higher frequency resulting in a distorted pulse. It is important to note that this is only the {\em radiated} pulse; the dispersive properties will double when the antenna is used for both transmit and receive.

\section{Dispersion Model}
Since the values of both $\alpha$ and $\tau$ remain constant for each cell in the sinuous antenna analyzed, the antenna is considered to be a log-periodic structure \cite{duhamel1987dual}. Thus, the radiated fields at frequency $\omega$ will repeat, since the structure repeats (scaled in size), at frequencies $\tau^n\omega$ where $n$ is an integer \cite{duhamel1957broadband}. A dispersion model for log-periodic antennas has been presented in the literature \cite{olvera2017dispersive, knop1970on, mclean2004the}. As will be demonstrated, this model may also be successfully applied to log-periodic sinuous antennas.

The model represents the phase due to dispersion as
\begin{equation}\label{eq:mod}
    \Phi^d_{mod}(\omega)=-\phi_0\ln\frac{\omega}{\omega_0},
\end{equation}
where $\phi_0=-\pi/\ln\tau$ \cite{mclean2004the}. The value $\omega_0$ controls the zero crossing of the phase model and is generally set to the highest frequency of operation (where the dispersion is defined to be zero); for this case, $\omega_0/2\pi=\text{10 GHz}$ was used. Fig. \ref{fig:phase} shows some discrepancy between the dispersion model and the simulation results. An optimization procedure\footnote{The optimization was done using MATLAB's global optimizer to find the best values for $\phi_0$ and $\omega_0$ when fitting the simulated phase $\Phi^d_{sim}$ starting with the initial suggested values.}, similar to what was done in \cite{olvera2017dispersive}, may be employed to produce an improved model $\Phi^d_{opt}$ as shown in Fig. \ref{fig:phase}. The default and optimized model parameters are compared in Table \ref{tbl:opt}. The group delay can then be computed from the phase model using (\ref{eq:gd}) and is shown in Fig. \ref{fig:gd}. As can be seen, the model fits the delay well with only a slight improvement obtained from the curve fit optimization. However, the model degrades at frequencies below 800 MHz where the sinuous antenna is no longer radiating as intended. A similar model developed for spiral antennas implemented a constant delay below the antenna's intended operating frequency \cite{mcfadden2009analysis}. Here, $\Phi_{opt}^d$ has a constant delay $\tau_c$ of 3.69 ns imposed for frequencies below 800 MHz.

\begin{table}[t]
\centering
\caption{Dispersion Model Parameters}
\label{tbl:opt}
\begin{tabular}{lccc}\toprule
 & \multicolumn{3}{c}{Model Parameters}  \\
 & $\phi_0$ (rad) & $\omega_0/2\pi$ (GHz) & $\tau_c$ (ns)\\ \midrule 
Default   & 20.01 & 10.0 & N/A \\
Optimized & 18.39 & 10.8 & 3.69 \\ \bottomrule 
\end{tabular}
\end{table}

The compressed radiated pulse $E_{comp}^x(r_p,t)$ was then computed using the optimized dispersion models as
\begin{multline}\label{eq:tf2}
    E_{comp}^x(r_p,t)=\\ \mathcal{F}^{-1}\biggl\{\mathcal{F}[v_{pulse}(t)]
    \frac{E^x_{sim}(r_p,\omega) \exp[-j\Phi^d_{opt}(\omega)]}{v_{sim}(\omega)}\biggr\},
\end{multline}
and is shown in Fig. \ref{fig:pulse}. For comparison, Fig. \ref{fig:pulse} also shows the compressed pulse computed with the default dispersion model. While not perfect, the correction causes the radiated pulse to closely match the shape of the input voltage with the optimized model giving the best result.

\section{GPR Simulations}
\begin{figure}
	\centering{\includegraphics[width=\columnwidth]{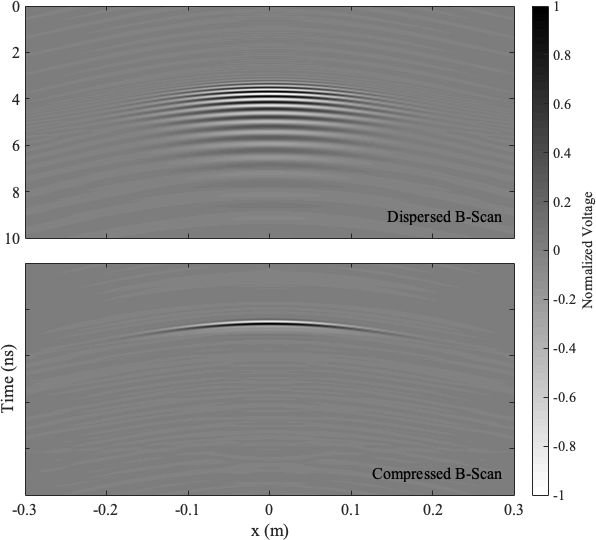}}
    \caption{GPR simulation results: dispersed B-scan (top), and compressed B-scan using the presented dispersion model (bottom).}
	\label{fig:bscan}
\end{figure}

The model was applied to a simulated GPR scenario where the sinuous antenna was scanned over a sandy-soil half-space as shown in Fig. \ref{fig:gprsim}. The model was simulated with the CST time-domain solver using the built in dispersive model for sandy soil. The radiated electric field was probed at a 20 cm depth and used to compute the returned signal from an electrically small linear scatter via the reciprocity model developed in \cite{mcfadden2009analysis}. The corresponding B-scans, both with and without pulse compression, are displayed in Fig.\ref{fig:bscan}. Only the default phase model with a fixed time delay was used to compress the returned pulse. The phase model was applied twice to the received voltage in order to compensate for dispersion produced during both transmit and receive. As can be seen, the model is able to successfully compress the pulse thereby significantly increasing the GPR's range resolution.


\definecolor{sand}{RGB}{237,201,175}
\begin{figure}
	\centering
	\begin{tikzpicture}[scale=1.0]
	\fill [white] (-4,0) rectangle (4,1);
	
	\draw[fill=black] (-0.75,0.4) rectangle (0.75,0.5);
	\node at (0,0.75) {Antenna};
	
    \draw[fill=sand] (-4,0) rectangle (4,-2);
	
    \draw[->, line width=0.3mm] (.9,0.45) -- ++(0.75,0) node[right] {$x$};
	\draw[->, line width=0.3mm] (0,0.25) -- ++(0,-0.6) node[below] {$z$};

    \node at (-2, 0.2) {2.5 cm};
    \draw[dashed] (-0.9,0.45) -- ++(-0.75,0);
    \draw[<-] (-1.5,0.5) -- ++(0,0.5);
    \draw[<-] (-1.5,-0.05) -- ++(0,-0.45);
    \draw[<->] (1,-0.05) -- (1,-1.45);
    \fill[black] (0,-1.5) circle (0.75mm);
    \draw[dashed] (0.15,-1.5) -- (1.1,-1.5);
    \node at (1.6,-0.75) {20 cm};
    \node at (-0.6,-1.3) {Target};

	\end{tikzpicture}
	\caption{Illustration of the GPR simulations: the sinuous antenna is scanned over a sandy-soil half-space.}
	\label{fig:gprsim}
\end{figure}
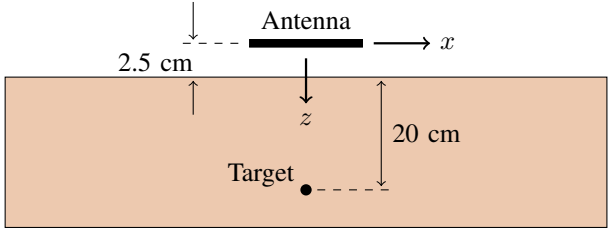

\section{Limitations of the Model}
The presented dispersion model is based on the assumption of log-periodic antenna operation. When the actual radiation from the antenna breaks this assumption the dispersion model becomes invalid. This was evident with the low-frequency dispersion discussed above. Another situation which reduces the effectiveness of the dispersion model is radiation from the bow-tie feed. A good rule is to keep $R_{in}<\lambda/4$, for the highest frequency desired, to prevent bow-tie radiation. Reducing $R_{in}$ also results in small trace widths at the feed which may be difficult to reliably manufacture. For this reason some have proposed breaking the log-periodic nature of the sinuous by letting $\tau$ vary with the radius \cite{sammeta2012quasi, sammeta2014improved}. In this case, the model would need to be updated since the antenna is now only quasi-log-periodic \cite{duhamel1987dual}.

Another potential pitfall is the unintended excitation of resonant modes which produce sharp variations in antenna gain over frequency \cite{kang2015modification}.
This may occur if the sinuous antenna design parameters and truncation method are not properly selected \cite{crocker2019on}.
The lower bound on the sinuous antenna operating frequency $\omega_L$ may be approximated as
\begin{equation}
    \omega_L =\frac{ 2\pi v}{4R_1(\alpha + \delta)},
\end{equation}
where $v$ is the wave velocity and $\alpha$ and $\delta$ are specified in radians \cite{duhamel1987dual}.
Such a relationship may encourage GPR antenna designers to choose larger values of $\alpha$ for lower operating frequencies.
However, large values of $\alpha$ have been shown to cause log-periodic resonances.
Further, the traditional truncation of sinuous antennas produces a sharp end that becomes resonate at low frequencies. 
These resonances reduce the ability of dispersion models to accurately compress radiated pulses.

\begin{figure}[!tbp]
	\centering
	\includegraphics[width=0.8\columnwidth]{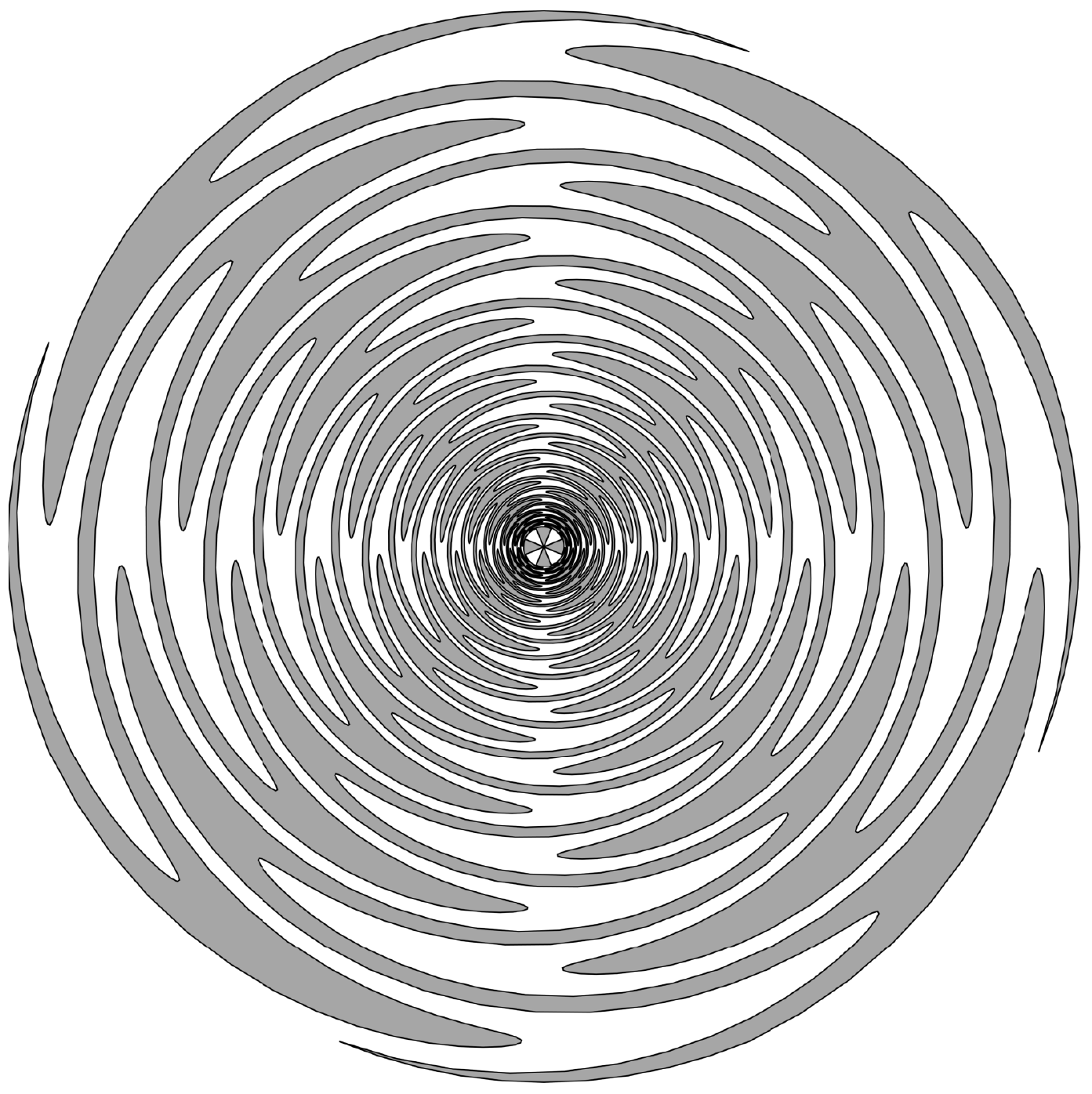}
    \caption{Traditional sinuous antenna having parameters: $N=\text{4}$ arms, $P=\text{20}$ cells, $R_1=\text{10}$ cm, $\tau=\text{0.8547}$, $\alpha=\text{65}^\circ$, and $\delta=\text{22.5}^\circ$. This antenna exhibits sharp discontinuities in the gain due to unintended resonate modes.}
	\label{fig:sinuous_res}
\end{figure}

\begin{figure}
	\centering{\includegraphics[width=\columnwidth]{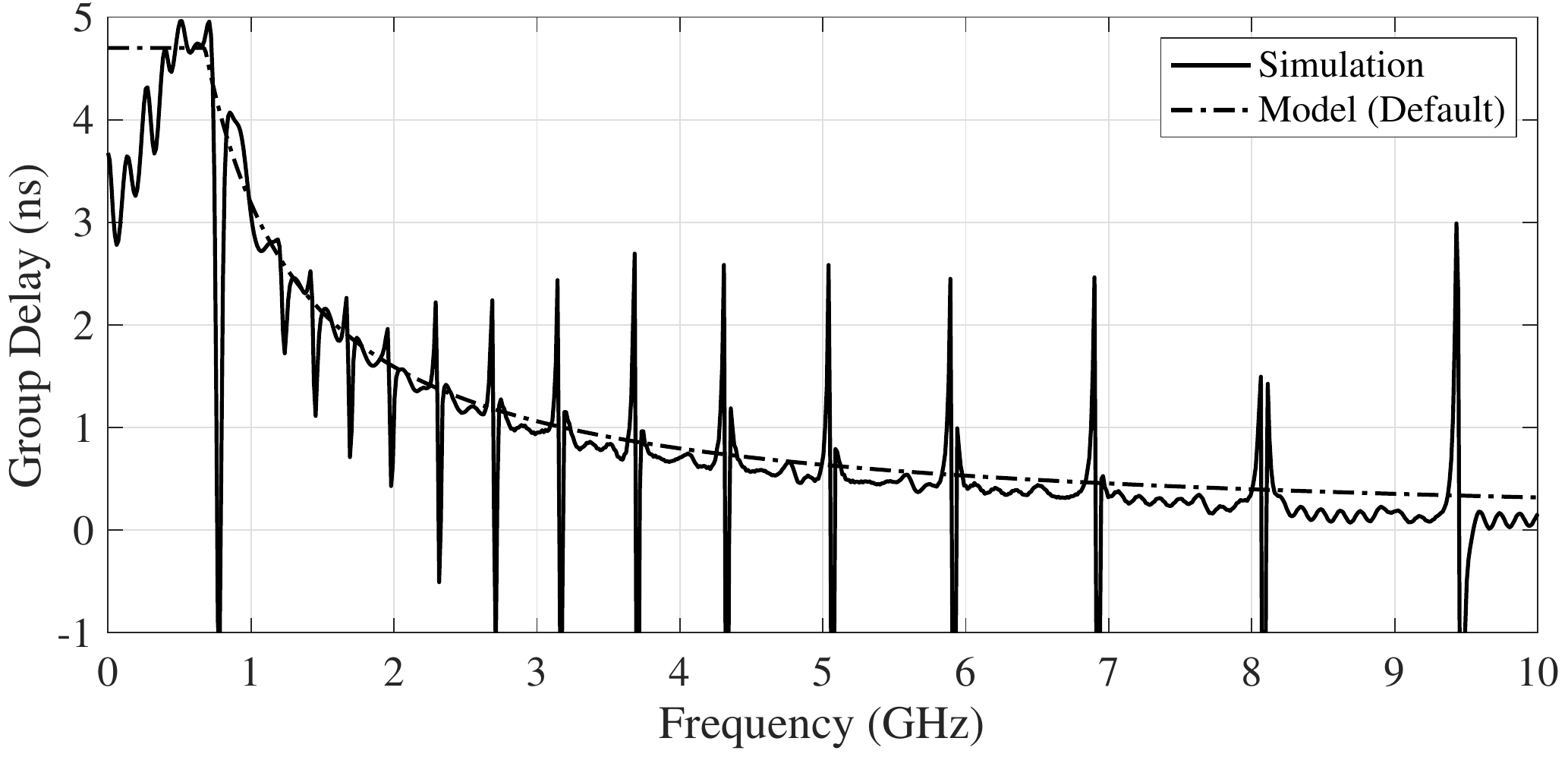}}
    \caption{Simulated and modeled (default with fixed delay cap) group delay of the radiation due to dispersion in the sinuous antenna with resonances.}
	\label{fig:gd_res}
\end{figure}

\begin{figure}
	\centering{\includegraphics[width=\columnwidth]{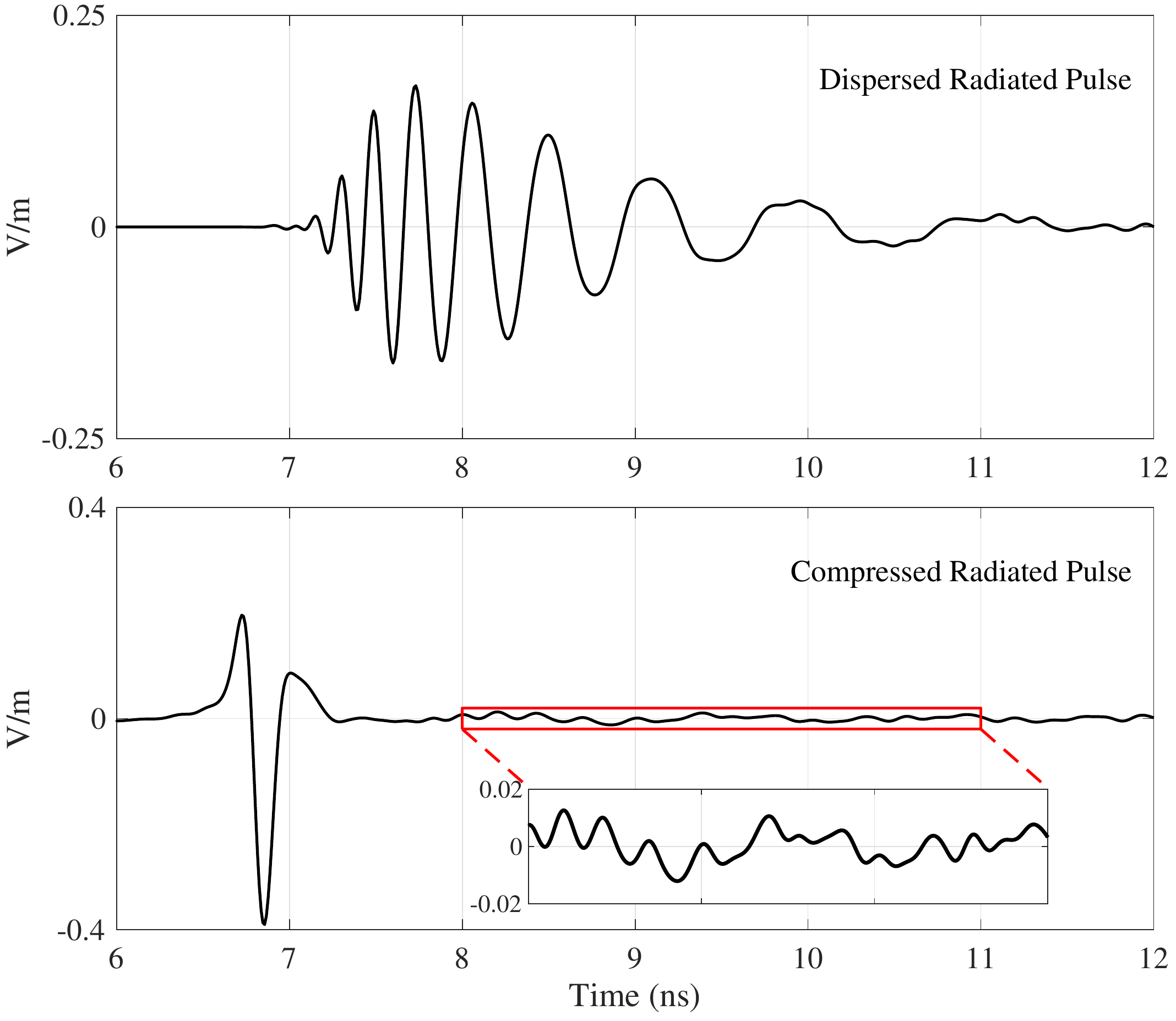}}
	\caption{Dispersed radiated pulse at 2 m (top), and the compressed radiated pulse at 2 m after the dispersion model has been applied (bottom). Note, the dispersion model does not compensated for the late-time ringing due to the gain variations.}
	\label{fig:pulse_res}
\end{figure}

In order to illustrate this, a traditionally truncated sinuous antenna with $\alpha=\text{65}^\circ$ (see Fig. \ref{fig:sinuous_res}) was simulated similarly to the antenna discussed previously. The group delay and corresponding dispersion model are shown in Fig. \ref{fig:gd_res}. 
Note that the default model with a fixed delay of 4.7 ns at low frequencies is used here since the sharp discontinuities prevent the optimization from finding a better curve fit.
The dispersion model is used to compress the radiated pulse as shown in Fig. \ref{fig:pulse_res}. As can be seen, the dispersion model is not able to compress the ringing due to the gain variations. Thus, the sinuous antenna must be designed in order to mitigate such ringing prior to the application of dispersion compensation techniques.

\section{Summary}
Sinuous antennas embody many characteristics that are advantageous to GPR applications. However, they are dispersive which reduces effectiveness when radiating UWB pulses. In this work, a model was presented for the compensation of dispersion in log-periodic sinuous antennas. The model is based on antenna design parameters and can be optimized for best fit. Such a model may have advantages over applying simulated or measured phase since the model is simplistic and can be adjusted in the field to accommodate antenna performance changes due to the environment.
Additionally, care must be taken when designing the sinuous antenna in order to ensure the applicability of such dispersion models.

\bibliographystyle{IEEEtran}
\bibliography{iwagpr2019_crocker_scott_sinuous_dispersion}

\end{document}